\documentstyle[12pt]{article}
\begin{document}
\topmargin -0.2cm \oddsidemargin -0.2cm \evensidemargin -1cm
\textheight 22cm \textwidth 12cm

\title{ A condition for the Bose-Einstein transition in the superfluid liquid helium.}  
\author{Minasyan V.N. \\
Yerevan, Armenia}

\date{\today}

\maketitle

\begin{abstract} 
First, a condition for the Bose-Einstein transition in the superfluid liquid helium is presented due to the creation of a free neutron spinless pairs in a liquid helium and a dilute neutron gas mixture. We proposed a new model of dilute Bose gas, where presence atoms in the condensate is a suppressor for the collective modes as well as a creator for single-particle excitations On other hand, it is shown that the terms, of the interaction between the excitations of the Bose gas and the density modes of the neutron, meditate an attractive interaction via the neutron modes, which in turn leads to a bound state on a spinless neutron pair. The lambda transition is defined by a condition for the Bose-Einstein transition, which transforms reflected neutron pair modes to single neutron modes.
\end{abstract} 

PACS $67.40.-w$

\vspace{100mm}

\vspace{5mm} 
 
{\bf 1. Introduction.} 
 
\vspace{5mm} 

The motivation for our theoretical study of the very low-temperature properties of the dilute hard sphere Bose gas is an attempt at a microscopic understanding of superfluidity in helium $^4$He. We proceed by discussing some experimental and theoretical investigations. 

The connection between the ideal Bose gas and superfluidity in helium was first made by London [1] in 1938. The ideal Bose gas undergoes a phase transition at sufficiently low temperatures to a condition in which the zero-momentum quantum state is occupied by a finite fraction of the atoms. This momentum-condensed phase was postulated by London to represent the superfluid component of liquid $^4$He. With this hypothesis, the beginnings of a two- fluid hydrodynamic model of superfluids was developed by Landau [2] where he predicted the notation of a collective excitations so- called phonons and rotons. 

The purely microscopic theory with mostly utilizes-technique, was first described by Bogoliubov [3] within the model of weakly non-ideal Bose-gas, with the inter-particle S- wave scattering. Based on the application of the presence of a macroscopic number of condensate atoms $N_0\approx N$ (where $N$ is the total number of atoms), Bogoliubov has dropped a density operator term which describes the fluctuation of atoms above the zero momentum level, the Bogoliubov obtained the dispersion curve for single particle Bogoliubov excitations (Bogoliubov phonon-roton modes).   

The dispersion curve of an excitations excited in superfluid helium has been accurately measured a function from momentum [4]. Within this experiment, in lambda transition, the position of a sharp peak inelastic neutron scattering intensity defines by energy of the single particle excitations, and there is appearing a broad component in inelastic neutron scattering intensity, at higher momenta of atoms. For explanation of the appearance of a broad component in inelastic neutron scattering intensity, the authors of papers [5-7] proposed to consider the presence of collective modes in the superfluid liquid $^4$He, which are represented a density excitations. In this respect, the authors of this letter predicted that the collective modes represent as the density quasiparticles [8]. In these works, these density excitations and density quasiparticles are appeared due to remained density operator term for describing atoms above the condensate, which was neglected by Bogoliubov [3]. 

The last time, the authors of letter [9] discovered an existence of scattering between atoms of the superfluid liquid helium, at lambda transition, which is confirmed by calculation of the dependence of the critical temperature on the interaction parameter as scattering length. On other hand, as it is indicated by authors of [4], there are two type excitations in superfluid helium, at lambda transition. These facts imply that it needs to revise the concept of determination condition for Bose-Einstein condensation in the superfluid liquid helium. Obviously, the peak inelastic neutron scattering intensity is connected with a registration of neutron modes by neutron-spectrometer, which in turn defines sort of excitations. Therefore, in this letter, we present to determine a condition for Bose condensation, at lambda transition, by a registration neutron-spectrometer a single neutron modes or neutron pair modes.

In this letter, we present a new model of a nonideal Bose gas for describing of the superfluid liquid helium. The given model is based on the application of the Penrose-Onsager definition of the Bose condensation [10] which is based on the condition for a condensed fraction of atoms  $\frac{N_0}{N}=const$. The later explains a broken of Bose-symmetry law for the atoms of the Bose gas in the condensate level, and gives a fully exact solution to the model of dilute non-ideal Bose gas, at quantity of the condensate fraction $0\leq\frac{N_0}{N}\leq 1$. In this context, the new model of a nonideal Bose gas, presented herein, leads to an absence of the collective modes because appearance of atoms in the condensate is the suppressor for collective modes as well as the creator for single-particle excitations. Therefore, we prove that the density excitations and the density quasiparticles, proposed by authors of letters [5-8], are unphysical. On other hand, at selection condition for condensed fraction $\frac{N_0}{N}\ll 1$, we may state that new model of a dilute Bose gas describes a thermodynamic property of  a realistic superfluid liquid helium.

Further, we investigate a helium liquid-dilute neutron gas mixture where exists the term of interaction between the Bogoliubov modes and the density modes of the neutrons, which due to application a canonical transformation of the Hamiltonian of system, the term of the interaction between the density of the Bogoliubov modes and the density of the neutron modes is removed by meditated an effective attractive interaction between the neutron modes, which in turn determines a bound state on neutron pair.

\vspace{5mm}
{\bf 2. New model of a dilute  Bose gas.}
\vspace{5mm}

For beginning, we present a new model of a dilute Bose gas for describing property of superfluid liquid helium. The given model considers a system of $N$ identical interacting atoms via S-wave scattering. These atoms, as spinless Bose-particles,  have a mass $m$ which are confined in a box of volume $V$. The main part of the Hamiltonian of such system is expressed in the second quantization form as: 

\begin{equation}
\hat{H}_a= \sum_{\vec{p}\not=0}\frac{p^2}{2m}
\hat{a}^{+}_{\vec{p}}\hat{a}_{\vec{p}}+
\frac{1}{2V}\sum_{\vec{p}\not=0}U_{\vec{p}}\hat{\varrho}_{\vec{p}}
\hat{\varrho}^{+}_{\vec{p}}
\end{equation}

Here  $\hat{a}^{+}_{\vec{p}}$ and $\hat{a}_{\vec{p}}$ are, respectively, the
"creation" and  "annihilation" operators of a free atoms with 
momentum $\vec{p}$;  $U_{\vec{p}}$ is the Fourier transform 
of a S-wave pseudopotential in the momentum space:

\begin{equation}
U_{\vec{p}} =\frac{4\pi d \hbar^2}{m }
\end{equation}

where $d$ is the scattering amplitude; and 
the Fourier component of the density operator presents as 

\begin{equation}
\hat{\varrho}_{\vec{p}}=\sum_{\vec{p}_1}
\hat{a}^{+}_{\vec{p}_1-\vec{p}}\hat{a}_{\vec{p}_1}
\end{equation}
  
According to the Bogoliubov's theory [3], it is a necessary to separate the atoms in the condensate from those atoms filling states above the condensate. In this respect, the operators $\hat{a}_0$ and  $\hat{a}^{+}_0$ are replaced by c-numbers  $\hat{a}_0=\hat{a}^{+}_0=\sqrt{N_0}$  within approximation the presence of a macroscopic number of condensate atoms $N_0\gg 1$. This assumption leads to a broken  of Bose-symmetry law for atoms occupying in the condensate state. To be refused a broken from Bose-symmetry law for bosons in the condensate, we apply the Penrose-Onsager's definition of the Bose condensation [10]:

\begin{equation}
\lim_{N_0, N\rightarrow\infty}\frac{N_0}{N}=const
\end{equation}

This reasoning is a very important factor for microscopic investigation of the model non-ideal Bose gas because the presence of a macroscopic number of atoms in the condensate means $N_0>>N_{\vec{p}\not =0}$  (where $ N_{\vec{p}\not =0}$ is the occupation number of  atoms with momentum  $\vec{p}\not =0$). In this respect, we may postulate a following approximation for an occupation number of  atoms with momentum  $\vec{p}$:    

\begin{equation}
\lim_{N_0\rightarrow\infty}\frac{N_{\vec{p}}}{N_0}=\delta_{\vec{p},0}
\end{equation}

The next step is to find the property of operators 
$\frac{\hat{a}^{+}_{\vec{p}_1-\vec{p}}}{\sqrt{N_0}}$, 
$\frac{\hat{a}_{\vec{p}_1-\vec{p}}}{\sqrt{N_0}}$ by applying (5). Obviously,

\begin{equation}
\lim_{N_0\rightarrow\infty}\frac{\hat{a}^{+}_{\vec{p}_1-\vec{p}}}{\sqrt{N_0}}=\delta_{\vec{p}_1, \vec{p}} 
\end{equation} 
and
\begin{equation}
\lim_{N_0\rightarrow\infty}\frac{\hat{a}_{\vec{p}_1-\vec{p}}}{\sqrt{N_0}}=
\delta_{\vec{p}_1, \vec{p}} 
\end{equation}

Excluding the term $\vec{p}_1=0$, :the density operators of bosons
$\hat{\varrho}_{\vec{p}}$ and $\hat{\varrho}^{+}_{\vec{p}}$ take the following forms:

\begin{equation}
\hat{\varrho}_{\vec{p}}= \sqrt{N_0}\biggl(\hat{a}^{+}_{-\vec{p}}+
\sqrt{2}\hat{c}_{\vec{p}}\biggl)
\end{equation}

and
\begin{equation}
\hat{\varrho}^{+}_{\vec{p}}=\sqrt{N_0}\biggl(\hat{a}_{-\vec{p}}+
\sqrt{2}\hat{c}^{+}_{\vec{p}}\biggl)
\end{equation}

where $\hat{c}_{\vec{p}}$ and $\hat{c}^{+}_{\vec{p}}$ are,
respectively, the Bose-operators of density-quasiparticles presented in reference [8] which in turn are the Bose-operators of bosons used in expressions (6) and (7):
:
\begin{equation}
\hat{c}_{\vec{p}}=\frac{1}{\sqrt{2N_0}} \sum_{\vec{p}_1\not=0}
\hat{a}^{+}_{\vec{p}_1-\vec{p}}\hat{a}_{\vec{p}_1}=\frac{1}{\sqrt{2}} \sum_{\vec{p}_1\not=0}\delta_{\vec{p}_1, \vec{p}} \hat{a}_{\vec{p}_1}=\frac{1}{\sqrt{2}}\hat{a}_{\vec{p}}
\end{equation}
and

\begin{equation}
\hat{c}^{+}_{\vec{p}} =\frac{1}{\sqrt{2N_0}}
\sum_{\vec{p}_1\not=0}\hat{a}^{+}_{\vec{p}_1}
\hat{a}_{\vec{p}_1-\vec{p}}=\frac{1}{\sqrt{2}} \sum_{\vec{p}_1\not=0}\delta_{\vec{p}_1, \vec{p}}\hat{a}^{+}_{\vec{p}_1}=
\frac{1}{\sqrt{2}}\hat{a}^{+}_{\vec{p}}
\end{equation}

Thus, we reach to the density operators of atoms $\hat{\varrho}_{\vec{p}}$ and 
$\hat{\varrho}^{+}_{\vec{p}}$, presented by Bogoliubov [3], at approximation $\frac{N_0}{N}=const$:

\begin{equation}
\hat{\varrho}_{\vec{p}}= \sqrt{N_0}\biggl(\hat{a}^{+}_{-\vec{p}}+
\hat{a}_{\vec{p}}\biggl)
\end{equation}

and
\begin{equation}
\hat{\varrho}^{+}_{\vec{p}}=\sqrt{N_0}\biggl(\hat{a}_{-\vec{p}}+
\hat{a}^{+}_{\vec{p}}\biggl)
\end{equation}

which displays that the density quasiparticles are absent.

The identical picture is observed in the case of the density excitations which was  predicted by the Glyde, Griffin and  Stirling  [5-7] where was proposed a presentation $\hat{\varrho}_{\vec{p}}$ in a following form: 

\begin{equation}
\hat{\varrho}_{\vec{p}}= \sqrt{N_0}\biggl(\hat{a}^{+}_{-\vec{p}}+
\hat{a}_{\vec{p}}+\tilde{\varrho}_{\vec{p}}\biggl)
\end{equation}
where terms involving $\vec{p}_1\not=0$ and $, \vec{p}_1\not=\vec{p}$  are 
written separately; and the operator $\tilde{\varrho}_{\vec{p}}$ describes  
the density-excitations: 

\begin{equation}
\tilde{\varrho}_{\vec{p}}=\frac{1}{\sqrt{N_0}} \sum_{\vec{p}_1\not=0, \vec{p}_1\not=\vec{p}}
\hat{a}^{+}_{\vec{p}_1-\vec{p}}\hat{a}_{\vec{p}_1}
\end{equation}

At inserting of (6) and (7) into (15), the term, representing as the density-excitations, vanishes because $\tilde{\varrho}_{\vec{p}}=0$.

Consequently, the Hamiltonian of system, presented in (1) by support of (12) and (13), reproduces an extension form of the Bogoliubov Hamiltonian, at approximation $\frac{N_0}{N}=const$:

\begin{equation}
\hat{H}_a = \sum_{\vec{p}\not=0}\biggl (\frac{p^2}{2m}+ mv^2\biggl) 
\hat{a}^{+}_{\vec{p}}\hat{a}_{\vec{p}}+
\frac{mv^2}{2}\sum_{\vec{p}\not=0} \biggl (\hat{a}^{+}_{-\vec{p}}\hat{a}^{+}_{\vec{p}}+\hat{a}_{\vec{p}}
\hat{a}_{-\vec{p}}\biggl)
\end{equation}

where $v=\sqrt{ \frac{ U_{\vec{p}} N_0}{mV}}=\sqrt{ \frac{4\pi d\hbar^2 N_0}{m^2V}}$ is the
velocity  of sound in the Bose gas which depends on the density atoms in the condensate $\frac{ N_0}{V}$.

For evolution of the energy level it is a necessary to diagonalize the Hamiltonian $\hat{H}_a$ which is accomplished by introduction of the Bose-operators $\hat{b}^{+}_{\vec{p}}$ and $\hat{b}_{\vec{p}}$ by using of the Bogoliubov linear transformation [3]:

\begin{equation} 
\hat{a}_{\vec{p}}=\frac{\hat{b}_{\vec{p}} + 
L_{\vec{p}}\hat{b}^{+}_{-\vec{p}}} {\sqrt{1-L^2_{\vec{p}}}} 
\end{equation}
where $L_{\vec{p}}$ is the unknown real symmetrical function
of  a momentum $\vec{p}$.

Substitution of (17) into (16) leads to 

\begin{equation}
\hat{H}_a=\sum_{\vec{p}}\varepsilon_{\vec{p}}\hat{b}^{+}_{\vec{p}}
\hat{b}_{\vec{p}}
\end{equation}

hence we infer that $\hat{b}^{+}_{\vec{p}}$ and
$\hat{b}_{\vec{p}}$ are the "creation" and "annihilation"
operators of a Bogoliubov quasiparticles with energy:

\begin{equation}
\varepsilon_{\vec{p}}=\biggl[\biggl(\frac{p^2}{2m}\biggl)^2 + p^2
v^2\biggl]^{1/2}
\end{equation}
 
In this context, the real symmetrical function $L_{\vec{p}}$
of  a momentum $\vec{p}$ is found

\begin{equation}
L^2_{\vec{p}}=\frac{\frac{p^2}{2m}+ mv^2-\varepsilon_{\vec{p}}}
{\frac{p^2}{2m}+
mv^2+\varepsilon_{\vec{p}}}
\end{equation}
 
It is well known, the strong interaction between the helium 
atoms is very important and reduces the condensate fraction 
to 10 percent or $\frac{N_0}{N}=0.1$  [4], at absolute zero. However, as we suggest a new model of dilute Bose gas, proposed herein, may have an significant application for describing of thermodynamic properties of the superfluid liquid helium because the S-wave scattering between two atoms, with coordinates $\vec{r}_1$ and $\vec{r}_2$ in the space of coordinate, is presented by the repulsive potential delta-function $U_{\vec{r}} =\frac{4\pi d \hbar^2\delta_{\vec{r}}}{m }$ from $\vec{r}=\vec{r}_1-\vec{r}_2$. On other hand, the presented model works on the condensed fraction $\frac{N_0}{N}\ll 1$ in differ from the Bogoliubov model where $\frac{N_0}{N}\approx 1$.
We can see that in statistical equilibrium, the equation for the density of charged bosons in the condensate is presented in a manner analogous to the one obtained by Bogoliubov [3]:

\begin{equation}
\frac{N_0 }{V}=\frac{N }{V}-
\frac{1}{V}\sum_{\vec{p}}\frac{L^2_{\vec{p}}}{1-L^2_{\vec{p}}}-
\frac{1}{V}\sum_{\vec{p} }\frac{1+L^2_{\vec{p}}}{1-L^2_{\vec{p}}}
\overline{\hat{b}^{+}_{\vec{p}}\hat{b}_{\vec{p}}}
\end{equation}
where $\overline{\hat{b}^{+}_{\vec{p}}\hat{b}_{\vec{p}}}$
is the average number of  the Bogoliubov quasiparticles with the momentum $\vec{p}$:

$$
\overline{\hat{b}^{+}_{\vec{p}}\hat{b}_{\vec{p}}}=\frac{1}{e^{\frac{\varepsilon_
{\vec{p}} }{kT}}-1}
$$

\vspace{5mm}
{\bf 3. The effective attractive potential interaction between neutron modes in a helium liquid-dilute neutron gas mixture.}
\vspace{5mm}
 
We now attempt to describe the thermodynamic property of a helium liquid-dilute neutron gas mixture.  In this context, we consider a neutron gas as an ideal Fermi gas consisting of  $n$ free neutrons with mass $m_n $ which interact with $N$ interacting atoms of a superfluid liquid helium. The helium-neutron mixture is confined in a box of volume $V$. The Hamiltonian of a considering system $\hat{H}_{a,n}$ consists of the term of the Hamiltonian of Bogoliubov excitations $\hat{H}_{a}$ in (18) and the term of the Hamiltonian of a dilute Fermi neutron gas as well as the term of interaction between the density of the Bogoliubov excitations and the density of the neutron modes: 
\begin{equation}
\hat{H}_{a,n}=\hat{H}_n+\sum_{\vec{p}}\varepsilon_{\vec{p}}\hat{b}^{+}_{\vec{p}}\hat{b}_{\vec{p}}+\frac{1}{2V}\sum_{\vec{p}\not=0}U_0 \hat{\varrho}_{\vec{p}}\hat{\varrho}_{-\vec{p},n }
\end{equation}
with
\begin{equation}
\hat{H}_n=\sum_{\vec{p},\sigma }\frac{p^2}{2m_n}
\hat{a}^{+}_{\vec{p},\sigma }\hat{a}_{\vec{p},\sigma} +\frac{1}{2V}\sum_{\vec{p}}U_1 \hat{\varrho}_{\vec{p},n }\hat{\varrho}_{-\vec{p},n }
\end{equation}

where  $\hat{a}^{+}_{\vec{p},\sigma}$ and 
$\hat{a}_{\vec{p},\sigma }$ are, respectively,  the operators of creation and 
annihilation for free neutron with momentum $\vec{p}$, by the value of its 
spin z-component $\sigma=^{+}_{-}\frac{1}{2}$; $U_0$ and $U_1$ are, respectively, the Fourier
transforms of the repulsive interaction between  the density of the Bogoliubov excitations and the density modes of the neutrons as well as between neutron modes: 
$$
U_0 =\frac{4\pi d_0\hbar^2}{\mu} 
$$
and
$$
U_1 =\frac{4\pi d_1\hbar^2}{m_n} 
$$

where $d_0$ is the scattering amplitude between a helium atoms and neutrons; $d_1$ is the scattering amplitude between neutron modes; $\mu =\frac{m\cdot m_n}{m+m_n}$ is the relative mass.

Hence, we note that the Fermi operators $\hat{a}^{+}_{\vec{p},\sigma}$ and $\hat{a}_{\vec{p},\sigma }$ satisfy to 
the Fermi commutation relations $[\cdot\cdot\cdot]_{+}$ as:

\begin{equation}
\biggl[\hat{a}_{\vec{p},\sigma}, \hat{a}^{+}_{\vec{p}^{'},
\sigma^{'}}\biggl]_{+} =
\delta_{\vec{p},\vec{p^{'}}}\cdot\delta_{\sigma,\sigma^{'}}
\end{equation}

\begin{equation}
[\hat{a}_{\vec{p},\sigma}, \hat{a}_{\vec{p^{'}}, \sigma^{'}}]_{+}= 0
\end{equation}

\begin{equation}
[\hat{a}^{+}_{\vec{p},\sigma}, \hat{a}^{+}_{\vec{p^{'}}, 
\sigma^{'}}]_{+}= 0
\end{equation}

The density operator of neutrons with spin $\sigma$ in momentum 
$\vec{p}$ is defined as
\begin{equation}
\hat{\varrho}_{\vec{p},n }=\sum_{\vec{p}_1, \sigma }
\hat{a}^{+}_{\vec{p}_1-\vec{p},\sigma }\hat{a}_{\vec{p}_1,\sigma }
\end{equation}
where $\hat{\varrho}^{+}_{\vec{p},n }=\hat{\varrho}_{-\vec{p},n }$

The operator of total number of neutrons is
$$
\sum_{\vec{p},\sigma}\hat{a}^{+}_{\vec{p},\sigma}\hat{a}_{\vec{p},\sigma}=
\hat{n}
$$
On other hand, the density operator, in the term of the Bogoliubov quasiparticles $\hat{\varrho}_{\vec{p}}$ included in (22), is expressed by following form, to application (17) into (12):

\begin{equation}
\hat{\varrho}_{\vec{p}}= \sqrt{N_0}\sqrt{\frac{1+ L_{\vec{p}}}{1- L_{\vec{p}}}}\biggl(\hat{b}^{+}_{-\vec{p}}+
\hat{b}_{\vec{p}}\biggl)
\end{equation}

Hence, we note that the Bose- operator  $\hat{b}_{\vec{p}}$ commutates with  the Fermi operator $\hat{a}_{\vec{p},\sigma }$ because the Bogoliubov excitations and neutrons are an independent.

Now, inserting of a value of operator $\hat{\varrho}_{\vec{p}}$ from (28) into (22), which in turn leads to reducing the Hamiltonian of system $\hat{H}_{a,n}$:

\begin{eqnarray}
\hat{H}_{a,n}&=&\sum_{\vec{p},\sigma }\frac{p^2}{2m_n}
\hat{a}^{+}_{\vec{p},\sigma }\hat{a}_{\vec{p},\sigma}+\frac{1}{2V}\sum_{\vec{p}}U_1 \hat{\varrho}_{\vec{p},n }\hat{\varrho}_{-\vec{p},n }+\nonumber\\
&+&\sum_{\vec{p}}\varepsilon_{\vec{p}}\hat{b}^{+}_{\vec{p}}\hat{b}_{\vec{p}}+\frac{ U_0\sqrt{N_0}}{2V}\sum_{\vec{p}} \sqrt{\frac{1+ L_{\vec{p}}}{1- L_{\vec{p}}}}\biggl(\hat{b}^{+}_{-\vec{p}}+\hat{b}_{\vec{p}}\biggl)\hat{\varrho}_{-\vec{p},n }
\end{eqnarray}

Hence, we note that the Hamiltonian of system $\hat{H}_{a,n}$ in (29) is a similar to the Hamiltonian of system an electron gas-phonon gas mixture which was proposed by the Fr$\ddot o$lich at solving of the problem superconductivity (please, see the Equation (16) in H. Fr$\ddot o$lich, Proc.Roy. Soc, {\bf A215}, 291-291 (1952) in  the reference [10] ), contains a subtle error in the term of the interaction between the density of phonon modes and the density of electron modes which represents a third term in right side of Equation (16) in [10] because the later is described by two sums, one from which goes by the wave vector $\vec{w}$ but other sum goes by the wave vector  $\vec{k}$. This fact contradicts to the definition of the density operator of the electron modes  $\hat{\varrho}_{\vec{w}}$ (please, see the Equation (12) in [11]) which in turn already contains the sum by the wave vector $\vec{k}$, and therefore, it is not a necessary to take into account so-called twice summations from $\vec{k}$ and $\vec{w}$ for describing of the term of the interaction between the density of phonon modes and the density of electron modes Thus, in the case of the Fr$\ddot o$lich, the sum must be taken only by wave vector $w$, due to definition of the density operator of electron modes with the momentum of phonon $\vec{w}$. 

To allocate anomalous term in the Hamiltonian of system $\hat{H}_{a,n}$,  which denotes by third term in right side in (29), we apply the Fr$\ddot o$lich approach [10] which allows to do a canonical transformation for
the operator $\hat{H}_{a,n}$ within introducing a new operator $\tilde{H}$:

\begin{equation}
\tilde{H}=\exp\biggl(\hat{S}^{+}\biggl)\hat{H}_{a,n}
\exp\biggl(\hat{S}\biggl)
\end{equation}

which is decayed by following terms: 

\begin{equation}
\tilde{H}=\exp\biggl(\hat{S}^{+}\biggl)\hat{H}_{a,n}
\exp\biggl(\hat{S}\biggl) = \hat{H}_{a,n}-[\hat{S},\hat{H}_{a,n}]+
\frac{1}{2}[\hat{S},[\hat{S},\hat{H}_{a,n}]]-\cdots
\end{equation}

where the operators represent as:
\begin{equation}
\hat{S}^{+}=\sum_{\vec{p}}\hat{S^{+}_{\vec{p}}}
\end{equation}
and   
\begin{equation}
\hat{S}=\sum_{\vec{p}}\hat{S_{\vec{p}}}
\end{equation}
and satisfy  to a condition $\hat{S}^{+} = -\hat{S}$

In this respect, we assume that

\begin{equation}
\hat{S_{\vec{p}}}=A_{\vec{p}}\biggl (\hat{\varrho}_{\vec{p},n}
\hat{b}_{\vec{p}}-
\hat{\varrho}^{+}_{\vec{p},n }\hat{b}^{+}_{\vec{p}}\biggl)
\end{equation}

where $A_{\vec{p}}$ is the unknown  real symmetrical function from  a momentum
$\vec{p}$. In this context, at application $\hat{S_{\vec{p}}}$ from (34) to (33) with taking into account $\hat{\varrho}^{+}_{-\vec{p},n }=\hat{\varrho}_{\vec{p},n }$, then we obtain 
\begin{equation}
\hat{S}=\sum_{\vec{p}}\hat{S_{\vec{p}}}= 
\sum_{\vec{p}}A_{\vec{p}}\hat{\varrho}_{\vec{p},n}\biggl (
\hat{b}_{=\vec{p}}-\hat{b}^{+}_{\vec{p}}\biggl)
\end{equation}
In analogy manner, at $\hat{\varrho}^{+}_{-\vec{p},n }=\hat{\varrho}_{\vec{p},n }$, we have
\begin{equation}
\hat{S}^{+}=\sum_{\vec{p}}\hat{S^{+}_{\vec{p}}}=
\sum_{\vec{p}}A_{\vec{p}}\hat{\varrho}^{+}_{\vec{p},n}\biggl (
\hat{b}^{+}_{\vec{p}}-\hat{b}_{-\vec{p}}\biggl)=
- \sum_{\vec{p}}A_{\vec{p}}\hat{\varrho}_{\vec{p},n}\biggl (
\hat{b}_{-\vec{p}}-\hat{b}^{+}_{\vec{p}}\biggl)
\end{equation}

Obviously, the given form of the operator $\hat{S}$ in (35) coincides with  a presentation of one, which was included by the Fr$\ddot o$lich  in Equation (18) by reference [10], within the function $\varphi (\vec{k},\vec{w})$ from momenta $\vec{k}$ and $\vec{w}$, which in turn represents as the c-number and equals the $\varphi (\vec{k},\vec{w})=\varphi (\vec{w})$, where $\varphi (\vec{w})$ is the real function from $\vec{w}$ which is a similar to the value $ A_{\vec{p}}$ proposed in this work. The later result is connected with a presentation of a correct form, for the term of interaction between the density phonon modes and the density of electrons, connected with the Fr$\ddot o$lich Hamiltonian in [10], as it is mentioned in above.  

To find  $A_{\vec{p}}$, we substitute  (29), (35) and (36) into  (31). Then,

\begin{equation}
[\hat{S},\hat{H_{a,n} }]=\frac{1}{V}\sum_{\vec{p}} 
A_{\vec{p}}U_0\sqrt{N_0}\sqrt{\frac{1+ L_{\vec{p}}}
{1- L_{\vec{p}}}}\hat{\varrho}_{\vec{p},n}
\hat{\varrho}_{-\vec{p},n}+
\sum_{\vec{p}} A_{\vec{p}} 
\varepsilon_{\vec{p}}\biggl(\hat{b}^{+}_{\vec{p}}+\hat{b}_{-\vec{p}}\biggl) \hat{\varrho}_{-\vec{p},n}
\end{equation}

\begin{equation}
\frac{1}{2}[\hat{S},[\hat{S},\hat{H}_{a,n}]]=
\sum_{\vec{p}} A^2_{\vec{p}} 
\varepsilon_{\vec{p}}\varrho_{\vec{p},n}
\hat{\varrho}_{-\vec{p},n}
\end{equation}

and $[\hat{S}, [\hat{S},[\hat{S},\hat{H}_{a,n}]]]=0$ within application a Bose commutation relations as 
$[\varrho_{\vec{p}_1,n},\hat{\varrho}_{\vec{p}_2,n}]=0$ and $[\hat{a}^{+}_{\vec{p}_1,\sigma}\hat{a}_{\vec{p}_1,\sigma},
\hat{\varrho}_{\vec{p}_2,n}]=0$. 

Thus, the form of new operator $\tilde{H}$ in (31) takes a following form:

\begin{eqnarray}
\tilde{H}& =&\sum_{\vec{p}}\varepsilon_{\vec{p}}
\hat{b}^{+}_{\vec{p}}\hat{b}_{\vec{p}}+
\frac{1}{2V}\sum_{\vec{p}}U_0\sqrt{N_0}
\sqrt{\frac{1+ L_{\vec{p}}}{1- L_{\vec{p}}}}
\biggl(\hat{b}^{+}_{-\vec{p}}+\hat{b}_{\vec{p}}\biggl)
\hat{\varrho}_{-\vec{p},n }-
\nonumber\\
&-&\frac{1}{V}\sum_{\vec{p}} A_{\vec{p}}U_0
\sqrt{N_0}\sqrt{\frac{1+ L_{\vec{p}}}
{1- L_{\vec{p}}}}\hat{\varrho}_{\vec{p},n }
\hat{\varrho}_{-\vec{p},n}-\sum_{\vec{p}} A_{\vec{p}} 
\varepsilon_{\vec{p}}
\biggl(\hat{b}^{+}_{-\vec{p}}+\hat{b}_{\vec{p}}\biggl) 
\hat{\varrho}_{-\vec{p},n }+
\nonumber\\
&+&\sum_{\vec{p}} A^2_{\vec{p}} 
\varepsilon_{\vec{p}}\hat{\varrho}_{\vec{p},n }
\hat{\varrho}_{-\vec{p},n}+\sum_{\vec{p},\sigma }\frac{p^2}{2m_n}
\hat{a}^{+}_{\vec{p},\sigma }
\hat{a}_{\vec{p},\sigma}+\frac{1}{2V}\sum_{\vec{p}}U_1 \hat{\varrho}_{\vec{p},n }\hat{\varrho}_{-\vec{p},n }
\end{eqnarray}

The transformation of the term of the interaction between the density of the Bogoliubov modes and the density neutron modes is made by removing of a second and fifth terms in right side of (39) which leads to obtaining of a quantity for
$A_{\vec{p}}$:

\begin{equation}
A_{\vec{p}}=\frac{U_0 \sqrt{N_0}}{2\varepsilon_{\vec{p}} V}\cdot\sqrt{\frac{1+ L_{\vec{p}}}{1- L_{\vec{p}}}}
\end{equation}

In this respect, we reach to reducing of the new Hamiltonian of system (39):

\begin{eqnarray}
\tilde{H}& =&\sum_{\vec{p}}\varepsilon_{\vec{p}}
\hat{b}^{+}_{\vec{p}}\hat{b}_{\vec{p}}+
\sum_{\vec{p},\sigma }\frac{p^2}{2m_n}
\hat{a}^{+}_{\vec{p},\sigma }\hat{a}_{\vec{p},\sigma}+\frac{1}{2V}\sum_{\vec{p}}U_1 \hat{\varrho}_{\vec{p},n }\hat{\varrho}_{-\vec{p},n }-\nonumber\\
&-&\frac{1}{V}\sum_{\vec{p}} A_{\vec{p}}U_0\sqrt{N_0}\sqrt{\frac{1+ L_{\vec{p}}}{1- L_{\vec{p}}}}\hat{\varrho}_{\vec{p},n }\hat{\varrho}_{-\vec{p},n}+\sum_{\vec{p}} A^2_{\vec{p}} \varepsilon_{\vec{p}}\hat{\varrho}_{\vec{p},n }\hat{\varrho}_{-\vec{p},n}
\end{eqnarray}

As result, the new form of Hamiltonian system takes a following fom:

\begin{equation}
\tilde{H}= \sum_{\vec{p}}\varepsilon_{\vec{p}}
\hat{b}^{+}_{\vec{p}}\hat{b}_{\vec{p}}+\hat{H}_n
\end{equation}
where $\hat{H}_n $ is the effective Hamiltonian of 
a neutron gas which contains an effective interaction between neutron modes:

\begin{equation}
\hat{H}_n =\sum_{\vec{p},\sigma }\frac{p^2}{2m_n}
\hat{a}^{+}_{\vec{p},\sigma }\hat{a}_{\vec{p},\sigma} +\frac{1}{2V}\sum_{\vec{p}}\biggl(V_{\vec{p}}+ U_1\biggl)
\hat{\varrho}_{\vec{p},n}\hat{\varrho}_{-\vec{p},n}
\end{equation}

where $V_{\vec{p}}$ is the effective potential of the interaction between neutron modes which takes a following form at substituting a value of $ A_{\vec{p}}$ from (40) into (41): 

\begin{equation}
V_{\vec{p}}= -2A_{\vec{p}}U_0\sqrt{N_0}\sqrt{\frac{1+ L_{\vec{p}}}{1- L_{\vec{p}}}}+ 2A^2_{\vec{p}} \varepsilon_{\vec{p}}V=-\frac{ U^2_0 N_0\biggl (1+ L_{\vec{p}}\biggl )}{ V \varepsilon_{\vec{p}} \biggl (1- L_{\vec{p}}\biggl )}
\end{equation}

In this letter, we consider following cases:
1. At low momenta atoms of a helium  $p<<2mv$, the Bogoliunov's quasiparticles in (19) represent as the phonons with energy $\varepsilon_{\vec{p}}\approx pv$  which in turn defines a value $L^2_{\vec{p}}\approx \frac{1-\frac{p}{mv}}{1+\frac{p}{mv}}\approx \biggl(1-\frac{p}{mv}\biggl)^2$ in (20)  or  $L_{\vec{p}}\approx 1-\frac{p}{mv}$. In this context, the effective potential between neutron modes takes a following form:

\begin{equation}
V_{\vec{p}}\approx -\frac{2 m U^2_0 N_0}{V p^2}= -\frac{4\pi \hbar^2 e^2_1}{p^2}
\end{equation}

The value $ e_1$ is the effective charge, at a small momenta of atoms: 

$$
e_1=\frac{U_0}{\hbar }\sqrt{\frac{ m N_0}{2V \pi }}
$$

2. At high momenta atoms of a helium $p>>2mv$, we obtain $\varepsilon_{\vec{p}}\approx \frac{p^2}{2m}+ mv^2$ in (19) which in turn defines $L_{\vec{p}}\approx 0$ in (20).  Then, the effective potential between neutron modes presents as:

\begin{equation}
V_{\vec{p}}\approx -\frac{ m U^2_0 N_0}{V p^2}= -\frac{4\pi \hbar^2 e^2_2}{p^2}
\end{equation}

where $e_2$ is the effective charge, at high momenta of atoms:
$$
e_2=\frac{U_0}{2\hbar }\sqrt{\frac{ m N_0}{V \pi }}
$$
  
Consequently, in both cases, a following form presents the effective scattering between two neutrons in momentum space:

\begin{equation}
V_{\vec{p}}= -\frac{4\pi \hbar^2 e^2_*}{p^2}
\end{equation}

where $ e_*= e_1$, at small momenta of atoms; and $ e_*= e_2$, at high momenta. 

\vspace{5mm}
{\bf 4. Creation Spinless Neutron Pairs and Single Neutrons.}
\vspace{5mm}
 
We now find the potential interaction between two neutrons in the coordinate space:

\begin{equation}
V (\vec{r})=\frac{1}{V}\sum_{\vec{p}}\biggl(V_{\vec{p}}+ U_1\biggl)\cdot e^{i\frac{\vec{p}\vec{r}}{\hbar}}=\frac{1}{V}\sum_{\vec{w}} \biggl(V_{\vec{w}}+ U_1\biggl)\cdot e^{i\vec{w}\vec{r}}
\end{equation}
where $V_{\vec{w}}=-\frac{4\pi  e^2_*}{w^2}$ is defined by wave number $w$.

Hence, we take into consideration a condition for S-wave scattering between neutron modes within using of a condition $w_f d_1\ll 1$ (where $w_f=\biggl(\frac{3\pi^2 n}{V}\biggl)^{\frac{1}{3}}$ is the Fermi wave number for neutron gas; $d_1$ is the scattering amplitude between neutrons). This reasoning implies that 
the scattering between two neutrons is presented in the coordinate space by a following form:

\begin{equation}
V (\vec{r})=\frac{1}{V}\sum_{\vec{w}} V_{\vec{w}}\cdot e^{i\vec{w}\vec{r}}=4\pi\int^{w_f}_{0} \biggl(V_{\vec{w}}+ U_1\biggl) w^2\frac{sin (w r)}{w r}d w
\end{equation}

where we introduce a following approximation as $ \frac{sin (w r)}{w r}\approx 1-\frac{w^2r^2}{6}$ because the conditions $w\leq w_f$ and $w_f d\ll 1$ lead to $ w_f r \ll 1$  ($w_f=\biggl(\frac{3\pi^2 n}{V}\biggl)^{\frac{1}{3}}$ is the Fermi wave number). The later condition defines a state for distance $r$ between two neighboring electrons $r\ll \frac{1}{w_f}=\biggl(\frac{V}{3\pi^2 n}\biggl)^{\frac{1}{3}}$. 

In this context, by taking into consideration $\frac{4\pi w^3_f}{3}=\frac{n}{2V}$, we obtain    

\begin{equation}
V (\vec{r})\approx -16\pi^2 e^2_* \biggl(\frac{3\pi^2 n}{V}\biggl)^{\frac{1}{3}}+ \frac{U_1 n}{2V} +\biggl (\frac{\pi e^2_*}{3}\frac{n}{V}-\frac{U_1}{2} \biggl(\frac{n}{V}\biggl)^{\frac{5}{3}}\biggl )\cdot r^2
\end{equation}

This approximation means that there is an appearance of a screening character in the effective scattering because as we see the later depends on the density electron modes. 

We now attempt to investigation of the effective Hamiltonian of a neutron gas in (43) is rewrite down in the space of coordinate by following form:

\begin{equation}
\hat{H}_n =\sum^{\frac{n}{2}}_{i=1}\hat{H}_i =-\frac{\hbar^2}{2m_n}\sum^{n}_{i=1} \Delta_I +\sum_{i<j} V (\mid\vec{r}_i-\vec{r}_j\mid )
\end{equation}

where $\hat{H}_i $ is the Hamiltonian of system consisting two neutrons with opposite spin which have a coordinates $\vec{r}_i $ and $\vec{r}_j $:
\begin{equation}
\hat{H}_i=-\frac{\hbar^2}{2m_n}\Delta_i-\frac{\hbar^2}{2m_n}\Delta_j + V (\mid\vec{r}_i-\vec{r}_j\mid )
\end{equation}
The transformation of considering coordinate system to the relative coordinate $\vec{r}=\vec{r}_i-\vec{r}_j $ and the coordinate of center mass $\vec{R}=\frac{\vec{r}_i+\vec{r}_j }{2}$, we have 

\begin{equation}
\hat{H}_i=-\frac{\hbar^2}{4m_n}\Delta_R-\frac{\hbar^2}{m_n}\Delta_r+ V (\vec{r})
\end{equation}

To find the binding energy $E<0$ of neutron pair, we search the solution of Schr$\ddot o$dinger equation with introduction of wave function $\psi(\vec{r})$:
$$
\hat{H}_i \psi_s (\vec{r})=E\psi_s (\vec{r})
$$
In this respect, we have a following equation 

\begin{equation}
-\frac{\hbar^2}{m_n}\Delta_r\psi_s (\vec{r})+ V (\vec{r})\psi_s (\vec{r}) = E\psi(\vec{r})
\end{equation}
which may determine the binding energy $E<0$ of neutron pair
Inserting value of $ V (\vec{r})$ from (50), and denoting $E=E_s$

\begin{eqnarray}
\biggl[-\frac{\hbar^2}{m_n}\Delta_{r}&-&16\pi^2 e^2_* \biggl(\frac{3\pi^2 n}{V}\biggl)^{\frac{1}{3}}+ 
\frac{U_1 n}{2V} -
\biggl (\frac{U_1}{2} \biggl(\frac{n}{V}\biggl)^{\frac{5}{3}} -
\nonumber\\
&-&\frac{\pi e^2_*}{3}\frac{n}{V} 
\biggl)\cdot r^2\biggl]
\psi_s (r)=E_s \psi_s (r)
\end{eqnarray}

We transform the form (55) by a following form:

\begin{equation}
\frac {d^2 \psi_s (r)}{d r^2} + \biggl (\lambda-\theta ^2 r^2 \biggl ) \psi_s (r)=0
\end{equation}

where we take 
$$
\theta=-\frac{\sqrt{ m_n }}{\hbar}\sqrt{ 
\frac{\pi e^2_*}{3}\frac{n}{V}-\frac{U_1}{2} \biggl(\frac{n}{V}\biggl)^{\frac{5}{3}} }
$$ 
and 
$$
\lambda =\frac{ m_n }{\hbar^2}\biggl(E_s +16\pi^2 e^2_* \biggl(\frac{3\pi^2 n}{V}\biggl)^{\frac{1}{3}}- 
\frac{U_1 n}{2V}\biggl)
$$

By application of the wave function 
$\psi_s (r)$ via the Chebishev-Hermit function 
$H_s (it)$ from an imaginary number as argument  $it$ [12] (where 
$i$ is the imaginary one; $t$ is the real number; $s=0;1;2;\cdots$), the equation (61) has a following solution as:

$$
\psi_s(\vec{r})=e^{-\theta\cdot r^2}H_s(\sqrt{\theta }\cdot r) 
$$
where 
$$
H_s(it)=i^s e^{-t^2}\frac{d^s e^{t^2}}{d t^s}
$$ 
at $\theta<0$ within
$$
\lambda=\theta (s+\frac{1}{2})
$$

Consequently, the quantity of the binding energy $E_s$ of neutron pair with mass $m_0=2m_n$ is rewritten as: 
\begin{equation}
E_s =-\frac{\hbar}{\sqrt{ m_n} }\sqrt{ 
\frac{\pi e^2_*}{3}\frac{n}{V}-\frac{U_1}{2} \biggl(\frac{n}{V}\biggl)^{\frac{5}{3}} }\biggl(s+\frac{1}{2}\biggl) -16\pi^2 e^2_* \biggl(\frac{3\pi^2 n}{V}\biggl)^{\frac{1}{3}}+
\frac{U_1 n}{2V}
< 0
\end{equation}
at $s=0;1;2;\cdots$

The normal state of neutron pair corresponds to quantity  $s=0$ which defines the maximal quantity of the binding energy of neutron pair: 
\begin{equation}
E_0 =-\frac{\hbar}{\sqrt{ m_n} }\sqrt{ 
\frac{\pi e^2_*}{3}\frac{n}{V}-\frac{U_1}{2} \biggl(\frac{n}{V}\biggl)^{\frac{5}{3}} }-16\pi^2 e^2_* \biggl(\frac{3\pi^2 n}{V}\biggl)^{\frac{1}{3}} + 
\frac{U_1 n}{2V} <0
\end{equation}
which implies that the creation of neutron pair is appeared by the following condition:

\begin{equation}
16\pi^2 e^2_* \biggl(\frac{3\pi^2 n}{V}\biggl)^{\frac{1}{3}} > 
\frac{U_1 n}{2V}
\end{equation}

Thus, the spinless neutron pair with mass $m_0=2m_n$ is created in a helium liquid-dilute neutron gas mixture, at condition (59), which determines a bound state on a neutron pair with binding energy (58), The later depends on the density of atoms in the condensate $\frac{N_0}{V}$, and therefore, may define the temperature of lambda transition $T_{\lambda}$, at condition 

\begin{equation}
\frac{\pi e^2_*}{3}>\frac{U_1}{2} 
\biggl(\frac{n}{V}\biggl)^{\frac{2}{3}} 
\end{equation}
 At low momenta atoms of a helium  $p<<2mv$,
\begin{equation}
\frac{N_{0, T_{\lambda}}}{V}> \frac{3U_1}{32\pi^2 U^2_0} \frac{\hbar^2}{m}\biggl(\frac{n}{V}\biggl)^{\frac{2}{3}}
\end{equation}
but at high momenta atoms of a helium  $p>>2mv$,
\begin{equation}
\frac{N_{0, T_{\lambda}}}{V}>  \frac{3U_1}{16\pi^2 U^2_0} \frac{\hbar^2}{m}\biggl(\frac{n}{V}\biggl)^{\frac{2}{3}}
\end{equation}

Thus, in the state of temperatures $0\leq T < T_{\lambda}$, there are a creation neutron pairs but at $T\geq T_{\lambda}$, the neutron pair is decayed on two free neutrons because the condition (59) is broken because the condition (59) takes a following form
\begin{equation}
16\pi^2 e^2_* \biggl(\frac{3\pi^2 n}{V}\biggl)^{\frac{1}{3}} \leq
\frac{U_1 n}{2V}
\end{equation}
which leads to positive meaning of $E_0\geq  0$.

In conclusion, we note that the new model of Bose gas, 
presented in this letter, might be useful for describing 
of the thermodynamic properties of a dilute gas of the Boson-Fermion mixtures confined in traps. Hence, we note that the correction form of  
the Fr$\ddot o$lich Hamiltonian system, for describing property of  a 
phonon gas-electron gas mixture in [13], may lead to creation spinless electron pairs, 
in addition of a phonon gas. The presented model of a helium liquid-dilute neutron gas mixture gives explanation of the presence of a broad component in inelastic neutron scattering intensity, at lambda transition [4] because there is an appearance single neutron modes, in addition to neutron pair modes, which in turn determine a superfluid phase for helium.

\newpage 
\begin{center} 
{\bf References} 
\end{center} 
 
\begin{enumerate} 

\item
F.~London~, Nature, ~{\bf 141},~643~(1938)
\item 
L.~Landau~, J. Phys.(USSR), ~{\bf 5},~77~(1941); Phys.(USSR),
~{\bf 11},~91~(1947).
\item 
N.N.~Bogoliubov~, Jour. of Phys.(USSR), ~{\bf 11},~23~(1947)
\item 
N.M.~Blagoveshchenskii~ et al.,Phys. Rev. B ~{\bf 50}, ~16550~(1994)
\item 
H.R.~Glyde~ and A.~Griffin~., Phys.Rev.Lett.~{\bf 65},~1454~(1990). 
\item 
W.G.~Stirling~,H.R.~Glyde~,  Phys.Rev.B. ~{\bf 41},~4224~(1990) 
\item 
H.R.~Glyde~, Phys.Rev.B. ~{\bf 45},~7321~(1992)
\item 
V.N.~Minasyan~ et.al~,~ Phys.Rev.Lett.~{\bf 90},~235301~(2003)
\item
K. ~Morawetz ~et al,, ~Phys. Rev. B~ 76~, ~075116~ (2007)
\item 
O. ~Penrose~ and L. ~Onsager~, ~Phys. Rev.,~{\bf 104}~, ~576~ (1956)
\item
H.~Fr$\ddot o$lich~, Proc.Roy. Soc, {\bf A215}, 291-291 (1952).
\item 
M.A.~Lavrentiev~, and B.V. ~Shabat ~ "Nauka", ~ Moscow,~, (1973)
\item 
~Minasyan~ V.N. ~ "Creation of Electron Spinless Pairs in the Superconductivity"
 arXiv:0903.0223 (2009)  
\end{enumerate} 
\end{document}